\documentstyle[emulateapj]{article}
\submitted{Astrophysical Journal, in press}

\def\mic{{\,\mu{\rm m}}}

\lefthead{AGUIRRE}
\righthead{INTERGALACTIC DUST AND TYPE IA SUPERNOVAE}

\begin{document}
\title{Intergalactic Dust and Observations of Type Ia Supernovae}
\author{Anthony Aguirre}
\affil{Department of Astronomy, Harvard University\\
60 Garden Street, Cambridge, MA 02138, USA\\
email: aaguirre@cfa.harvard.edu}
 
\begin{abstract}
  
  Estimates of the cosmic star formation rate and of cluster
  metallicities independently imply that at $z \la 0.5$ the gas in the
  universe has substantial average metallicity: $1/10 \la Z/Z_\odot
  \la 1/3$ for $\Omega_{gas} = 0.05$.  This metal density probably
  cannot be contained in known solar-metallicity galaxies of density
  parameter $\Omega_* \approx 0.004$, implying significant enrichment
  of the intergalactic medium (IGM) by ejection of metals and dust
  from galaxies via winds, in mergers or in dust efflux driven by
  radiation pressure.  Galaxies have a dust/metal ratio of $\sim 0.5$
  in their interstellar media, but some fraction $(1-f) > 0$ of this
  must be destroyed in the IGM or during the ejection process.
  Assuming the Draine \& Lee dust model and preferential destruction
  of small grains (as destruction by sputtering would provide), I
  calculate the reddening and extinction of a uniform cosmological
  dust component in terms of $f$ and the minimum grain size $a_{min}$.
  Very small grains provide most of the reddening but less than half of
  the opacity for optical extinction. For $f \ga 0.3$ and $a_{min} \ga
  0.1\mic$, the intergalactic dust would be too grey to have been
  detected by its reddening, yet dense enough to be cosmologically
  important: it could account for the recently observed type Ia
  supernova dimming at $z \sim 0.5$ without cosmic acceleration. It
  would also have implications for galaxy counts and evolutionary
  studies, and would contribute significantly to the cosmic infrared
  background (CIB).  The importance of grey intergalactic dust of the
  described type can be tested by observations of $z=0.5$ supernovae
  in (rest) $R-$band or longer wavelengths and by the fluxes of a
  large sample of supernovae at $z > 1.$

\vskip0.1in
\noindent {\em Subject headings:} cosmology: observations, distance scale -- intergalactic medium --
dust, extinction
 
\end{abstract}

\section{Introduction}

The importance of dust extinction in the Galaxy has been recognized
since early in this century when star-counting surveys revealed
absorption of optical light by `dark clouds' (Barnard 1919).  It is
fortunate that extinction correlates relatively well with reddening in
the Galaxy, because it is difficult enough to accurately measure
either the distance to a typical astronomical source or its intrinsic
luminosity -- let alone both.  But knowing the intrinsic color
$(B-V)_i$ (using an unobscured line of sight) along with the observed
$B-V$ color and a reddening-extinction law $A_V =
R_V[(B-V)-(B-V)_i] \equiv R_VE(B-V)$, one can correctly
determine the distance of an object from its distance modulus, or vice
versa.  The value of $R_V$ varies markedly within the Milky Way ($ 3
\la R_V \la 6$; Mathis 1990) and among different galaxies ($1.5 \la
R_V \la 7.2$; Falco et al.  1999), but a value of $R_V \simeq 3.2$ is
useful for many estimates of Galactic extinction.

Radiation from extragalactic objects is subject to extinction by dust
both inside and outside of the Galaxy.  However, while extragalactic
dust has received attention, our knowledge of its amount and properties is
rather limited, because methods useful for estimating dust density in
galaxies have proved less effective when extended to intergalactic
space.  For example, a number of groups have attempted to measure
extinction by dust in clusters using background-object counting, and
several claims of intracluster dust (Bogart \& Wagoner 1973; Boyle,
Fong \& Shanks 1988; Romani \& Maoz 1992) and extragalactic dust cloud
detections (Wsozlek et al. 1988 and references therein) have been
made, but even now these claims remain controversial (see Maoz 1995).
The strongest proposed limits on a diffuse distribution of
intergalactic dust with a Galactic reddening law have come from
studies of the redshift evolution of the mean quasar spectral index
(e.g. Wright 1981; Wright \& Malkan 1988; Cheng, Gaskell \& Koratkar
1991). These studies limit uniform dust of constant comoving density
to have $A_V(z=1) \la 0.05$ mag (from Wright \& Malkan 1988), and are
most sensitive to dust at $z > 1.$

While our knowledge of it is poor, intergalactic dust could have great
cosmological importance, as it could affect results concerning the
cosmic microwave (CMB) and cosmic infrared (CIB) backgrounds, galaxy
and quasar numbers at high $z$, galaxy evolution, large-scale
structure, etc.  This paper discusses intergalactic dust chiefly in
the context of its importance in measurements of the cosmological
deceleration parameter -- a subject discussed numerous times, first by
Eigenson (1949) and most recently in Aguirre 1999~(A99).

Conditions in the diffuse intergalactic medium (IGM) strongly disfavor
dust formation, so whatever intergalactic dust exists is probably
either the remnant of an early Population III epoch, or is formed in
galaxies and removed by some mechanism. (The remaining possibility,
that a substantial dust-forming population of extragalactic stars
exists, is not considered here.)  Previous investigations of
intergalactic dust have almost invariably assumed that it has
properties similar to that of Galactic dust; but this assumption is
not well justified.  Even among galaxies, $R_V$ varies by a factor of
four, and intergalactic dust may have creation, destruction, and
selection mechanisms quite different from dust in galaxies.  As argued
in A99, radiation pressure ejects grains with high opacity and a broad
opacity curve more efficiently than other grain types.  In
\S\ref{sec-dustdest} I discuss results suggesting that small grains
are preferentially destroyed by sputtering, both in the halos of
galaxies (during the ejection process) and perhaps in the IGM.

The large grains, while having significant mass, give very small
$E(B-V)$ reddening, and actually have higher visual opacity (per unit
mass) than dust which includes small grains.  Intergalactic dust of
this type would not have been detected by quasar reddening surveys.
Current data, described in \S\ref{sec-metals}, suggests that the
universe has been enriched to $\ga 1/10$ solar metallicity before
$z\sim 0.5$.  Section~\ref{sec-snae} shows that if a significant
fraction of this metal is locked in dust that is distributed fairly
uniformly, the dust extinction to $z\sim 0.5$ can explain the observed
progressive dimming of type Ia supernovae (Riess et al. 1998;
Perlmutter et al. 1999 [P99]) without cosmic acceleration.

The last section outlines ways in which the influence of the type of
intergalactic dust described here can be tested, probably most cleanly
by future supernova observations.

\section{Cosmic Metallicity}

\begin{figure}
\plotone{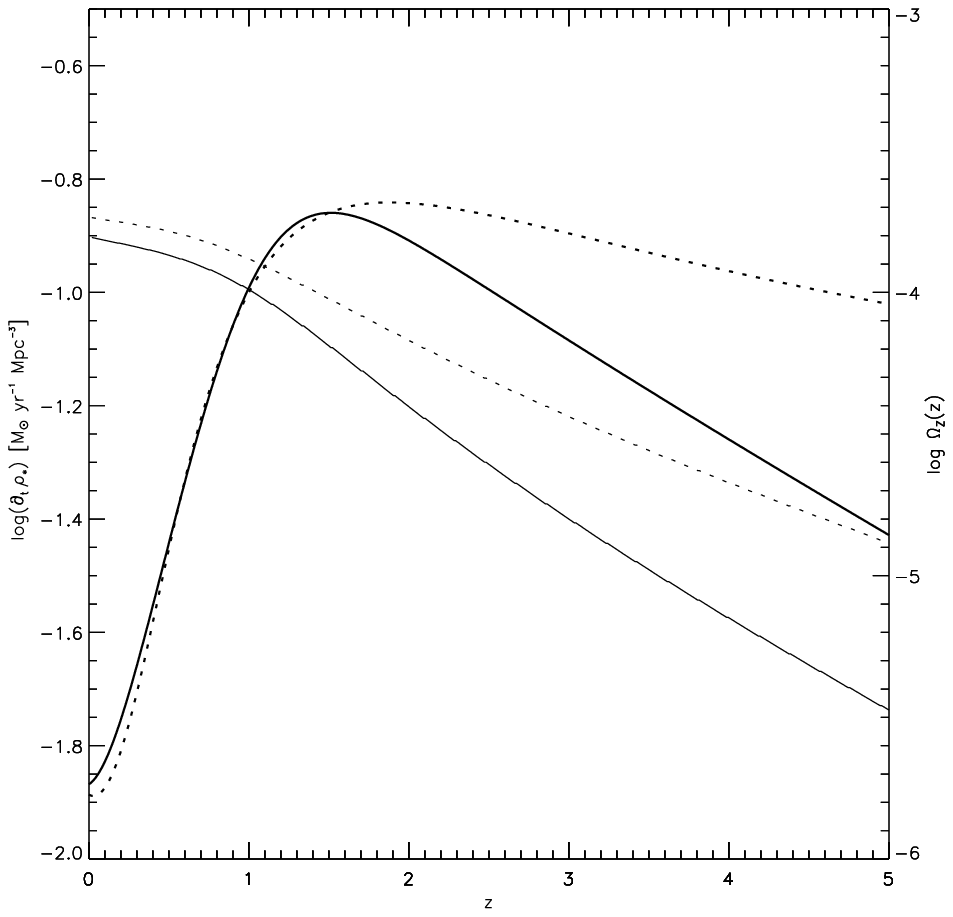}
\caption{Comoving star formation rate (thick; left axis) and integrated 
(from $z=10$) 
  cosmic metal density (thin; right axis), from Madau~(1999; solid)
  and Steidel et al.~(1999; dashed).  Adjusted for $H_0=65\,{\rm
    km\,s^{-1}\,Mpc^{-1}}$, $\Omega=0.2$.}
\label{fig-sfr}
\end{figure}

Although still very incomplete, our understanding of the evolution of
cosmic metallicity has improved dramatically during the last few
years.  Recent surveys (see e.g. Madau 1999 for a review) of galaxies
in the Hubble Deep Field show that the comoving star formation rate
(SFR) rises with redshift by a factor of ten to $z\sim 1-1.5$, past
which it either declines or levels off.  Using the observed star
formation rate and an assumed ratio of metal formation to star
formation of 1/42 (Madau et al. 1996), the total cosmic metallicity
(neglecting any population III contribution) may be estimated by
integrating over time the metal formation rate.  Figure~\ref{fig-sfr}
shows the result of this integration starting at $z=10$, using curves
from Madau (1999) and Steidel et al. (1999).  These demonstrate that
whether or not the SFR declines for $z \ga 1.5$, the current metal
density is $\Omega_Z \sim (1.5-2)h_{65}^{-2}\times 10^{-4}.$ The
estimates shown for $\Omega_Z(z)$ agree with other results in the
literature using similar methods: Pettini (1999) estimates
$\Omega_Z(z\simeq 2.5) \sim 6\times 10^{-5}$; Cen \& Ostriker (1999b)
calculate $\Omega_Z(z\simeq2.5) \sim 2\times 10^{-5}$ and
$\Omega_Z(z\simeq0.5)\simeq 1.1\times10^{-4}$.


Very interestingly, these estimates coincide with the `fossil
evidence' presented by Renzini (1997; 1998).  He argues that clusters
are essentially closed systems which contain all of the metal produced
by their stellar populations.  Stars of approximately solar
metallicity comprise only a fraction $\Upsilon_{cl} \approx
0.09h_{65}^{3/2}$ of the total cluster gas mass, yet the remaining
intracluster gas has $\approx 1/3$ solar metallicity.  The associated
metal production per unit of stellar mass can be written $M_Z \approx
M_*[1+3.15h_{65}^{-3/2}]Z_\odot$ (Renzini 1997).  Unless stars in
clusters produce metals much more efficiently than those in field
galaxies, this figure should apply to the $\Omega_* \approx 0.004$
(Fukugita, Hogan \& Peebles 1996) of stars in the universe, giving
$\Omega_{Z} \approx 3.1 \times 10^{-4}$.  Moreover, if the cosmic gas
density is $\Omega_{gas} \approx 0.05$, then the star formation
efficiency of clusters well represents that of the universe,
$\Upsilon_{IGM} \equiv {\Omega_* \over \Omega_{gas}} \approx
\Upsilon_{cl}$, further indicating that clusters are a fair sample.

Two obvious ratios link these numbers to estimates of intergalactic
dust density: the fraction $F_I$ of all metal residing in the IGM, and
the fraction $d_m$ of the intergalactic metal locked in dust.  At
high $z$, the primary data bearing upon these ratios are observations
of metal lines in quasar absorption spectra.  Pettini et al. (1997)
have measured the depletions of Cr and Zn in damped Lyman-$\alpha$
systems which Pei, Fall \& Hauser (1998) interpret as giving a dust/metal ratio
of $\approx 0.5$ for all $z \la 3$, similar to that of known galaxies.
This suggests that even at high $z$, source regions for the cosmic
metallicity have $d_m \approx 0.5$.  Whether this applies to metal as
it reaches the IGM depends on the degree to which the ejection process
destroys dust, or selectively ejects either gaseous metal or dust (to
be discussed in~\S\ref{sec-dustrem}).

Observations of lower column density Ly-$\alpha$ absorbers give
information about the gaseous metals in relatively cool regions of the
IGM.  Under some (important) assumptions about ionization corrections,
these line strengths indicates that areas with column density $N(H I)
\approx 10^{14-15}\,{\rm cm^{-2}}$ have $\sim 0.3\%$ solar
metallicity, giving $\Omega_{Z}^{ly}(z\sim3)\sim3\times 10^{-6}$ in
gaseous metal if this enrichment is uniform.  Direct integration
(effectively treating $N(H I) \la 10^{14}\,{\rm cm^{-2}}$ regions as
pristine) gives $\Omega_Z^{ly}(z\sim3) \ga 3\times 10^{-7}$ (Songaila
1997).  In light of the higher numbers derived from the SFR, the
assumption that most of the cosmic metallicity resides in the
Ly-$\alpha$ forest therefore leads to a `missing metals' problem, as
noted by Pettini (1999) and Renzini (1999), who argue that this
discrepancy may indicate that most metals are located in hot halos
around galaxies, in proto-clusters, or in a phase of the IGM hotter
than that which the Ly-$\alpha$ forest studies probe. Cen \& Ostriker
(1999a) reach the same conclusion on separate grounds, arguing on the
basis of both simple physical arguments and their numerical
simulations that the bulk of universal baryons at low $z$ should be
hot.

The idea that the IGM is fairly metal-rich gains more support from
arguments that hold at low redshift. If enriched to solar metallicity,
the current mass of stars and gas in known galaxies is sufficient to
contain the metals expected at $z\sim 2.5$ (if most galaxies had
already formed by then), but this does not hold for later epochs: as
also argued by Renzini (1998), the cosmic metallicity at $z=0$ derived
from cluster enrichment or by integration of the SFR cannot be
stored in currently observed galaxies, which can hold at most
\begin{equation}
\Omega_Z = 8\times 10^{-5}\left({Z_{gal}\over
    0.02}\right)\left({\Omega_{gal}\over 0.004}\right)
\end{equation}
The remaining $\approx 50-75 \%$ must exist in the intergalactic gas,
in extended halos, or hidden in undetected galaxies (which seems
unlikely).  This is exactly what one would expect on the basis of the
metal distribution in clusters: Renzini (1997) argues that while
intracluster gas contains about three times as much metal as the
cluster galaxies, there seems to be no compelling reason to expect
that metal ejection is much more efficient in clusters than in the
field\footnote{Even if ram-pressure stripping and/or mergers made
  clusters more efficient, this would predict that field galaxies
  would be about four times as metal-rich as cluster galaxies,
  contrary to observations. Also, clusters with higher velocity
  dispersion should have higher metallicity; Renzini finds no evidence
  for this.}, so most metals created from field galaxies should also
lie outside them.

To avoid the conclusion that the universe has substantial
intergalactic metallicity, it therefore seems that one would have to
argue both that cluster galaxies eject metals more efficiently
than field galaxies and have a different IMF, and also accept
that estimates of the SFR and/or the density of metal in galaxies are
incorrect by at least a factor of two.

In summary, the density of intergalactic dust can be estimated at
$z\sim 0.5$ as
\begin{equation}
\Omega_{dust}^{igm}(z\sim0.5) = F_I\times d_m \times \chi \times
10^{-4},
\label{eq-metdens1}
\end{equation}
with likely values of $1.5 \la \chi \la 3$ and $d_m \simeq 0.5$. The
argument that stars and gas in known galaxies cannot contain these
metals gives $0.5 \la F_I \la 0.75$; the larger number also
corresponds to the value derived by assuming that field galaxies eject
metal as cluster galaxies do.  The resulting estimate does not
take into account grain destruction.
\label{sec-metals}

\section{Removal of dust and metal from galaxies}

The conclusion that the IGM is fairly metal rich implies that metal
can be efficiently removed from the galaxies in which it forms.
Several ways in which galaxies can eject dust and metallic gas have
been studied in the context of clusters (for exactly the same reason),
but it rather unclear which mechanism is dominant.  The density of
intracluster gas is high enough that ram-pressure stripping of
galactic gas may be efficient (e.g. Fukumoto \& Ikeuchi 1996; Gunn \&
Gott 1972) in clusters, but it would be much less effective for a
galaxy in the general IGM.  The removal of gas by supernova-driven
winds has been widely discussed, and detailed simulations have been
performed investigating this effect in large starburst galaxies
(Suchkov et al.  1994) and in dwarves (Mac Low \& Ferrara 1998).  It
is widely thought that galactic winds also remove gas from ellipticals
(e.g. David, Forman \& Jones 1990).  Gnedin (1998) has claimed that
mergers provide the dominant metal removal mechanism, at least for $z
\ga 4.$ Finally, dust removal (without metallic gas) by radiation
pressure can be fairly efficient even for present-day spirals;
starburst galaxies with higher luminosities would be correspondingly
more effective.  In this section I will concentrate on radiation
pressure as the dominant dust expulsion mechanism because the effects
on the dust have been investigated most carefully in that scenario;
but I will discuss the other mechanisms briefly.

The dynamical evolution of a dust grain in a spiral galaxy is governed
primarily by the radiation pressure, gravity, the viscous gas drag and the
magnetic Lorentz force.  Starting with Chiao \& Wickramasinghe (1972), several
groups have studied these forces acting on grains in model galaxies
with some assumptions about the mass, gas, and luminosity
distributions.
Confirming the results of Chiao \& Wickramasinghe, Ferrara et al. (1990) find
that graphite grains of most sizes can escape most spiral galaxies,
and that silicate grains are marginally confined (though silicate grains
with $a\ga 0.05\mic$ can escape high luminosity spirals.)  

Calculating the grain dynamics for two specific galactic models (of
the Milky way and NGC 3198) in more detail, Ferrara et al. (1991) find
that in the Milky Way, silicate grains of radius $a = 0.1\mic$ have
typical speeds of $\sim 200-600\,{\rm km\,s^{-1}}$ and reach halo
radii $\sim 100$ kpc in $\sim 150-500$ Myr. Graphite grains of $a =
0.05\mic$ can move approximately twice as fast.  Another
investigation, by Shustov \& Vibe (1995), gives similar results. They
find that grains of size $0.07\mic \le a \le 0.2\mic$ are ejected most
effectively.  Silicate (graphite) grains of $0.1\mic$ starting 1~kpc
above the galactic disk attain speeds of $1000\,{\rm km\,s^{-1}}$
($2000\,{\rm km\,s^{-1}}$) and reach 100 kpc in 100 (40) Myr.  Shustov
\& Vibe stress that only dust starting somewhat above the disk can
escape, but this does not imply that dust ejection is inefficient: in
their model, the dust expulsion rate is up to $0.4M_\odot\,{\rm
  yr^{-1}}$, which in a Hubble time exceeds the entire metal content
of the Galaxy and gives $\Omega_{dust}^{igm}$ of order $10^{-4}$, even
assuming a constant SFR.  Most recently, Davies et al.  (1998) have
performed numerical calculations of dust removal by radiation
pressure, taking into account the opacity of the disk.  Disk opacity
reduces the radiation pressure at high galactic latitude, so Davies et
al. find dust expulsion less efficient than the previous
studies.\footnote{Of course a higher assumed intrinsic luminosity of
  the galaxy would cancel this effect; Davies et al.  only investigate
  one mass-to-light ratio.} Nevertheless, Davies et al.  predict the
removal of (at least) $0.1\mic$ graphite grains from their model
galaxy if there is fairly low disk opacity.  Smaller grains are
expelled much less efficiently.  Notably, Davies et al.  still
estimate that up to $90\%$ of the dust formed in spirals may be
ejected, and even calculate an estimate of $\sim 1$ mag of
intergalactic extinction across a Hubble distance.

All studies of radiation pressure driven dust removal have noted the
importance of magnetic fields on grain dynamics but have (effectively)
neglected them in their calculations, with the following
justifications given: 
\begin{itemize}
\item{Magnetic field lines are only sometimes
parallel to the disk, and only on large scales (all studies).}
\item{Grains are charged only part of the time (Ferrara et al. 1991; Davies
et al. 1998).}
\item{Radiation pressure can enhance Parker (1972) instabilities
that can lead to open field configurations with lines perpendicular to
the disk. (Chiao \& Wickramasinghe 1972; Ferrara et al. 1991).}
\end{itemize}
An additional justification is empirical: dust is actually observed
outside the disks of galaxies (Howk \& Savage 1999; Alton, Davies \&
Bianchi 1999; Davies et al. 1998 and references therein; Ferrara et
al. 1999 and references therein).  However the dust escapes the disk,
its presence proves that while magnetic fields are potentially
important and currently impossible to model in detail, they cannot be
perfectly effective at dust confinement. On the other hand, this does
not prove that dust can fully decouple from the gas.

While dust expelled by radiation pressure could not carry a
significant gas mass with it, other metal expulsion mechanisms
probably remove dust along with gas.  Alton et. al. (1999) have
presented observations of dust outflows in three nearby starburst
galaxies, concluding that the `superwinds' driving these outflows can
impart near-escape velocity on the dust, and that up to $10\%$ of the
dust mass of these galaxies could be lost in the observed outflows
alone.  These results lend observational support to the numerical
simulations of Suchkov et al. (1994) which predicted such outflows,
and further demonstrate that dust can escape along with metallic gas.
Lehnert \& Heckman (1996) have estimated the efficiency of metal
removal by winds in starburst galaxies using a large sample, and find
that galaxies could enrich the IGM to $\Omega_Z^{igm} \sim 5 \times
10^{-5}$.  This figure assumes a constant SFR and the authors estimate
that the enrichment is likely to be ten times higher with a more
realistic SFR.

Supernova-driven winds might also eject
dust from ellipticals.  Vereshchagin, Smirnov \& Tutukov (1989)
estimate that the ratio of galactic wind force to gravitational force
on a grain of radius $a$ is
$$F_W/F_G = {3\alpha \over 16\pi G}{V_W/a},$$
where $V_W$ is the wind
velocity and $\alpha$ is the specific mass ejection rate for the wind
in the galaxy.  For $\alpha \sim 10^{-19}\,{\rm s^{-1}}$ (applicable
for the present epoch; Vereshchagin et al. 1989; David et al. 1990)
and relatively slow winds ($V_W \sim 10-60\,{\rm km\, s^{-1}}$;
Vereshchagin et al. 1989), only very small grains escape; but for
starburst ellipticals, David et al. find rates of at least $5\times
(10^{-18}-10^{-17})\,{\rm s^{-1}}$ for the first $10^8\,{\rm yr}$ of
starburst activity, implying that the winds dominate gravity in grain
dynamics for grains up to $\sim 0.05-2.5\mic$, even for winds too
slow to escape the galaxy themselves.  Essentially the same argument
would apply to grains subject to winds in spirals, so whether dust
lies in relatively cool outflows or is exposed to the wind itself, it
is difficult to see how it could avoid being driven out with the
metallic gas.

Gnedin (1998) has performed high resolution
cosmological simulations that suggest that mergers are the dominant
metal removal mechanism, at least at high ($z \ga 4$) redshift.  This
mechanism would eject gas with roughly the same dust/gas ratio as the
source galaxy.  

This brief survey of metal ejection mechanisms suggests that it is
difficult to efficiently remove metallic gas from galaxies without
also removing dust (although the converse of this would not be true if
radiation pressure is the dominant ejection mechanism).  Rough
estimates of the ejection efficiencies show that metal ejection rates
sufficient to account for the enrichment of the IGM estimated in
\S\ref{sec-metals} are reasonable.  While dust probably accompanies
gas as it leaves galaxies (or leaves by itself), studies of clusters
show that intracluster gas is not dust-rich.  The next section
addresses the probable cause of this disagreement: grain destruction
during the ejection process and in the IGM.

\label{sec-dustrem}

\section{Destruction of small grains}

A key result of the investigations by both Ferrara et al. (1991) and
Shustov \& Vibe (1995) is that grain destruction due to sputtering by
hot halo gas is relatively insignificant for grains of $a \sim
0.1\mic$ but very effective for grains with $a\sim 0.01\mic.$ While
$0.1\mic$ grains lose only $\sim 0.005\mic$ in radius, the small
grains are completely destroyed on a timescale of $\sim 500$ Myr.  The
sharp difference arises because for sputtering at fixed gas
temperature and grain velocity the destruction timescale $(1/a)(da/dt)
\propto 1/a$ (Draine \& Salpeter 1979a), and because small grains are
more affected by gas drag yet less propelled by radiation pressure,
hence move more slowly through the halo.\footnote{Size effects can be
  even stronger; Draine \& Salpeter (1979b) find that the most
  efficient dust destruction, sputtering in the `inter-cloud medium',
  is $\sim 500\,\times$ more effective in $0.01\mic$ grains than for
  $a=0.1\mic.$} Shustov \& Vibe conclude from their calculations that
the grains escaping intact from galaxies will have sizes $0.03\mic \la
a \la 0.2\mic$ for graphite particles and $0.07\mic \la a \la 0.2\mic$
for silicate particles.\footnote{These numbers are rather approximate
  because the authors computed results only for six grain radii.}
These conclusions depend on assumptions about the density,
temperature, and extent of the galactic halos, but the fact that both
groups obtain similar results suggests that the minimal grain size
surviving expulsion is probably of order $a_{min} \sim 0.05\mic.$

The efficiency of dust destruction in other metal removal processes
has not been calculated in detail and is difficult to estimate.  Dust
driven out by winds would be vulnerable to sputtering by the halo gas
as well as by the faster moving wind, though it will be somewhat
shielded if embedded in cool clumps of gas.  To estimate the effect of
the wind, let us assume a mass loss rate $\dot M$ (in solar masses/yr)
due to a wind leaving the galaxy radially with velocity $V_W$.  The
effect of this wind would be similar to the effect of a hot gas of
temperature $T_W \equiv m_p V_W^2/2k$ and (proton) number density $n_p
\sim 3\dot M / 16\pi R^2 V_W m_p$.  For $125\,{\rm km\,s^{-1}} \la v
\la 4000\,{\rm km\,s^{-1}}$ and $R \ga 10$ kpc, this gives $10^6\,{\rm
  K} \la T \la 10^9$ K and
\begin{equation}
n_p \la 1.9 \times 10^{-4}\,\dot M
\left({V_W \over 125\,{\rm
      km\,s^{-1}}}\right)^{-1}\,{\rm cm^{-3}}.
\end{equation}
Using Draine \& Salpeter's (1979a) sputtering rate for graphite
in this temperature range,
this corresponds to a lifetime of
$$
\tau_W \ga (7-16) \times 10^7\,\dot M^{-1} \left({V_W
  \over 125\,{\rm km\,s^{-1}}}\right)\left({a\over0.01\mic}\right)\,{\rm yr}.
$$
This is comparable to the ejection timescale, so this sputtering
could be important but is unlikely to completely destroy the dust.

Grain-grain collisions provide another important dust destruction
mechanism in galaxies and might be important in the early stages of
the ejection process.  For example, grain-grain collisions in
supernova shocks can efficiently shatter large grains into smaller
ones (e.g. Jones, Tielens \& Hollenbach 1996), so if supernova blowout
removes dust, there is a danger that shocks from the same supernovae
might shatter the large grains before the dust is expelled.  Shocks
may also play an important role in mergers.  On the other hand, it is
unclear whether the dust observed in the ISM is representative of
dust which has just formed, or already been shock-processed, or some
steady-state between the two.  Specifically, there is evidence for the
formation of large grains in novae (Shore et al. 1994), and possibly
in supernovae (see Wooden 1997 and Pun et al.  1995), and grains are
presumably larger in molecular clouds where high values of $R_V$ are
measured.  It may be, then, that pre-shock grains tend to be somewhat
larger, and the MRN distribution is more characteristic of grains
after significant shattering has occurred.  The assumption of this
paper is that dust leaving its progenitor galaxy will have an grain
size distribution characteristic of dust in the ISM.  In the absence
of significant shattering, this is probably conservative, since a
significant fraction of dust is contained in dense clouds with high
$R_V$ (e.g. Kim, Martin \& Hendry 1994).

\label{sec-dustdest}

\subsection{Dust Destruction in the IGM}

Rather little is known about the destruction of dust in the IGM.
Schmidt (1974) estimates that soft cosmic rays would provide the most
efficient destruction, but cannot determine whether or not the
destruction time would exceed the Hubble time; moreover, Draine \&
Salpeter (1979b) find that cosmic rays are unimportant dust destroyers
in the Galaxy (where they should be at least as effective as in the
IGM).  The hot gas component of the IGM, however, could sputter grains
effectively, even at low density.  Using again Draine \& Salpeter's
(1979a) estimate\footnote{More recent calculations by Tielens et al.
  give similar sputtering rates for carbon at $T \ga 10^7{\rm\,K}$, while their rates
  are somewhat higher for silicates and somewhat lower in both
  materials for $10^6{\,\rm K} < T < 10^7\,$K.}  , the lifetime can be
written
\begin{equation}
\tau \approx (3.5-9)\times10^9\,\left({a\over0.01\mic}\right)
h_{65}^{-2}\Omega_{gas}^{-1}
\delta^{-1}(1+z)^{-3}\,{\rm yr},
\end{equation}
where $\Omega_{gas}$ signifies the hot gas density in critical units and
$\delta$ is a clumping factor.  The Hubble time (for $\Omega=1$) is
$H^{-1}(z) = 1.6 \times 10^{10}h_{65}^{-1}(1+z)^{-3/2}\,{\rm yr}$,
suggesting the efficient destruction of grains for which
\begin{equation}
Q \equiv 0.14\left({a\over0.01\mic}\right)^{-1}\left({\Omega_{gas}\over
    0.05}\right)h_{65}\delta(1+z)^{3/2} \gg 1.
\end{equation}
The clumping factor (i.e.  the overdensity felt by a `typical' grain)
is quite uncertain, but the simulations of Cen \& Ostriker (1999b),
which numerically track the distribution of metallicity, indicate that
at $z\sim 0.5$, the mean universal metallicity approaches the
metallicity of $\delta \sim 100$ regions.  Regions of much higher
overdensity do not have much higher metallicity and hence cannot
contain most of the metals -- for example, $\delta \sim 1000$ regions
have only about twice the metallicity, so dense `subregions' can
contain only about 20\% of the metals in $\delta \sim 100$ regions.
If a `typical' grain experiences $\delta \sim 100$, then $Q \sim 26$
for $0.01 \mic$ grains and $Q \sim 2.6$ for $0.1\mic $ grains.  This is
suggestive (but {only} suggestive) that sputtering by hot
intergalactic gas might provide yet another mechanism by which grains
of $a \la 0.1\mic$ might be selectively destroyed.

Finally, note that the low mean dust density in the IGM and in
extended galaxy halos would strongly suppress the grain-grain
collisions thought to shatter large grains into small ones in the
galaxy\footnote{Equation~\ref{eq-galopt} gives the dust optical depth
  through the halo of a galaxy. The `optical depth' to an emerging
  grain would be of similar magnitude, so grain-grain collisions are
  probably unimportant unless high-$z$ galaxies are all heavily
  obscured by dust in their halos.}; since dust formation is also
inefficient in the IGM there is probably no source of {new} small
grains outside of galaxies.  

The efficiency of dust destruction depends in a rather complicated way
on the environment; moreover the type and details of the dominant
mechanism of metal ejection for galaxies are uncertain.  Thus the
arguments of this section are intended merely to make plausible the
chief {assumption} of this paper, which is that grains of size $a \la
0.05-0.1 \mic$ are removed (either by destruction or by failure to
escape their progenitor galaxies) from the grain-size distribution
characterizing dust outside of galaxies, whereas larger grains are
not.

\section{Density of Surviving Intergalactic Dust}

The estimate of the density of intergalactic dust in section
\S\ref{sec-metals} did not take into account dust destruction or the
preferential expulsion of dust.  Lets us assume that a mass fraction
$(1-f_{esc})$ of dust is destroyed as it leaves the disk and/or
traverses the halo, and that a further fraction $(1-f_{igm})$ is
destroyed in the IGM after the dust escapes the halos but before $z
\sim 0.5$.  There are three general scenarios indicated by the dust
ejection and destruction mechanisms outline above:

\begin{enumerate}
\item{Dust and gas leave together, with the galactic dust/metals ratio
    of $\approx 0.5$.  A fraction $f_{esc}f_{igm}$ of this survives,
    so that $d_m \approx 0.5 f_{esc}f_{igm}$. This scenario predicts a
    high enrichment of the IGM near galaxies and perhaps a substantial
    density of metal in galactic halos.}
\item{Dust, driven by radiation pressure, decouples from the gas
    either in the disk or in the inner halo, but is partially
    destroyed.  The gaseous metal (both destroyed dust and metal which
    escapes the disk but then decouples from dust) could (a) remain in
    the halo or could (b) return to the disk to form more dust,
    repeating the process.  In the former case galactic halos may be
    highly enriched and $d_m \approx 0.5 f_{esc}f_{igm}$; in the
    latter (unlikely) case galaxies should be very deficient in metals
    which are easily incorporated into dust, and $d_m \sim f_{igm}$ is
    possible.}
\item{Gaseous metals leave disks but dust remains ($d_m \ll 1$).
    While unlikely, this would lead to a highly enriched IGM and/or
    halo gas component but little intergalactic obscuration (like the
    case $f_{esc}f_{igm} \ll 1$).  Disks would be heavily enriched
    with elements that {do} form dust.}
\end{enumerate}

Assuming that $0.5 \la F_I \la 0.75$, equation~\ref{eq-metdens1} gives
\begin{equation}
\Omega_{dust}^{igm}(z\sim0.5)
\sim (4-11) \times10^{-5}f_{esc}f_{igm}
\label{eq-metdens2}
\end{equation}
for scenarios 1 and 2a.  The dust density would be higher by a factor
of up to $\sim 2/f_{esc}$ for scenario 2b, and very small for 3.

Since small grains are preferentially destroyed, but probably cannot
be created in halos and in the IGM, $f_{esc}f_{igm}$ effectively
determines the minimal grain size $a_{min}$.\footnote{Sputtering will
  also change the upper grain-size cutoff, in effect shifting the
  whole distribution toward smaller radii.  The neglect of this effect
  may be justified by the excess of large grains over the MRN
  prediction indicated by estimates of the actual grain-size
  distribution (see \S\ref{sec-othermodels}), and because sputtering
  may be more selective in destroying small grains than the $1/a$
  behavior would imply.}  The next section discusses the dependence of
dust properties on $a_{min}$, and gives corresponding values of
$f_{esc}f_{igm}$.
\label{sec-dustdens}

\section{Properties of the dust}

The absence of $a \la 0.1\mic$ grains would have important
implications for the properties of intergalactic dust.  The most
commonly used model for Galactic dust is the two component Draine \&
Lee (1984; DL) model consisting of silicate and graphite spheres with
a distribution in radius (as proposed by Mathis, Rumpl and Nordsieck
1977; MRN) of $N(a)da \propto a^{-3.5},\ 0.005\mic \le a \le
0.25\mic$.  After synthesizing dielectric functions for both graphite
and `astronomical silicate' and assuming a silicate/graphite mass
ratio $\sim 1$, DL demonstrated that the resulting model fits both the
observed opacity and polarization over a wide wavelength range
($0.1\mic \la \lambda \la 1000\mic$), most notably fitting the
observed features at $0.2175\mic$ and $10\mic$.  This paper employs
the DL model not because it is most likely to be correct, but because
there is little agreement as to what the correct grain model might be.  The
DL model is widely used and familiar, and hopefully (but by no means
certainly) captures the essential features of the dust.  Other models
are discussed briefly below.

An interesting aspect of the MRN distribution is that while the
geometrical cross section ($\propto a^2$) is dominated by small-radius
grains, the mass ($\propto a^3$) is dominated by grains of large
radii.  Thus, removing the very small grains can affect the opacity
curve dramatically, without radically changing the total dust mass.

\begin{figure}
\plotone{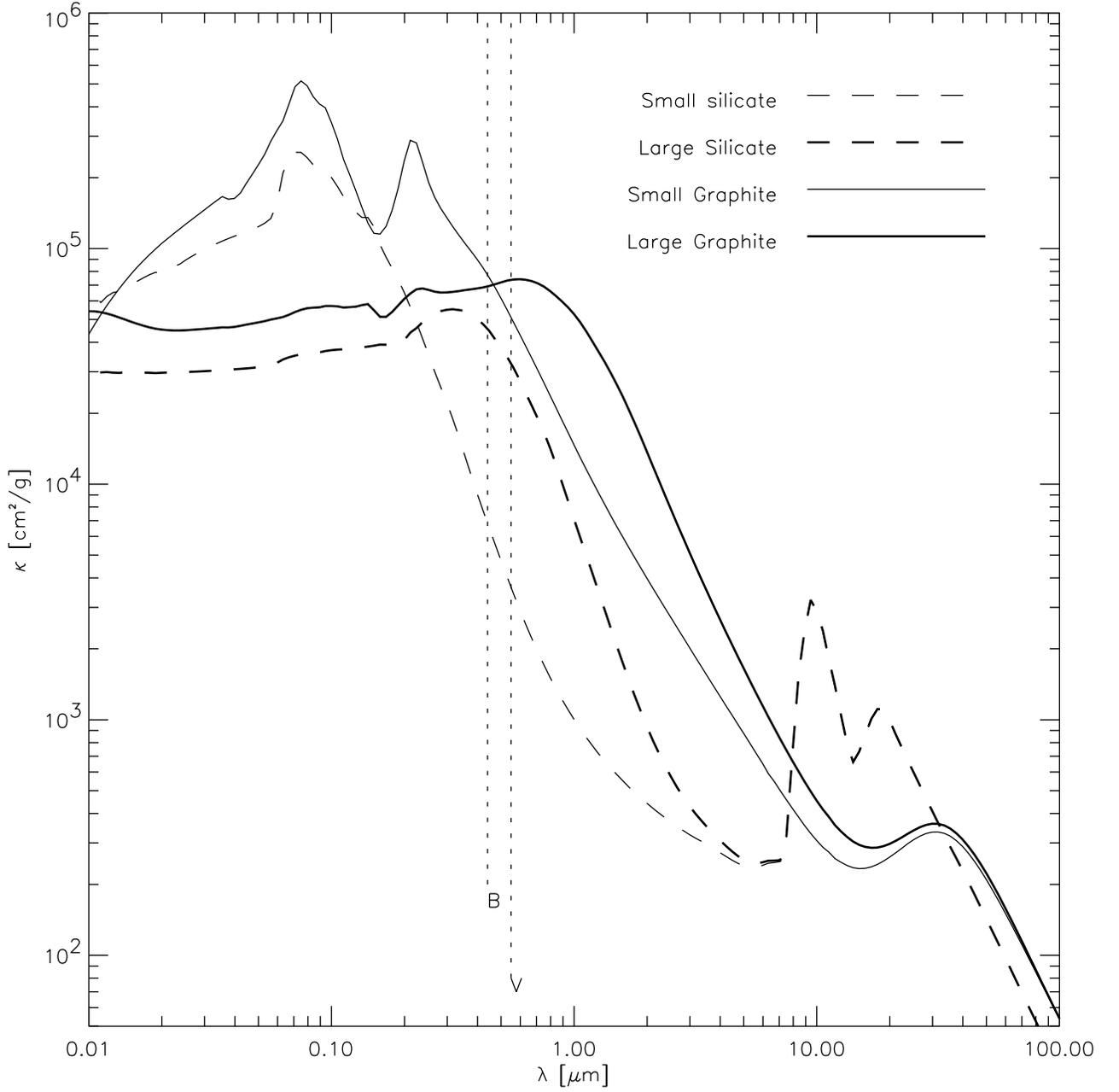}
\caption{Opacity (absorption+scattering) of Draine \& Lee model with truncated MRN distribution
  $a_{min}=0.005\mic$, $a_{max}=0.1$ for graphite (solid, thin) and
  silicate (dashed, thin), and for $a_{min}=0.1\mic$, $a_{max}=0.25$ for
  graphite (solid, thick) and silicate (dashed, thick).  Vertical
  dotted lines indicate $B$ and $V$-band centers.}
\label{fig-opacs}

\end{figure}
\begin{figure}
\plotone{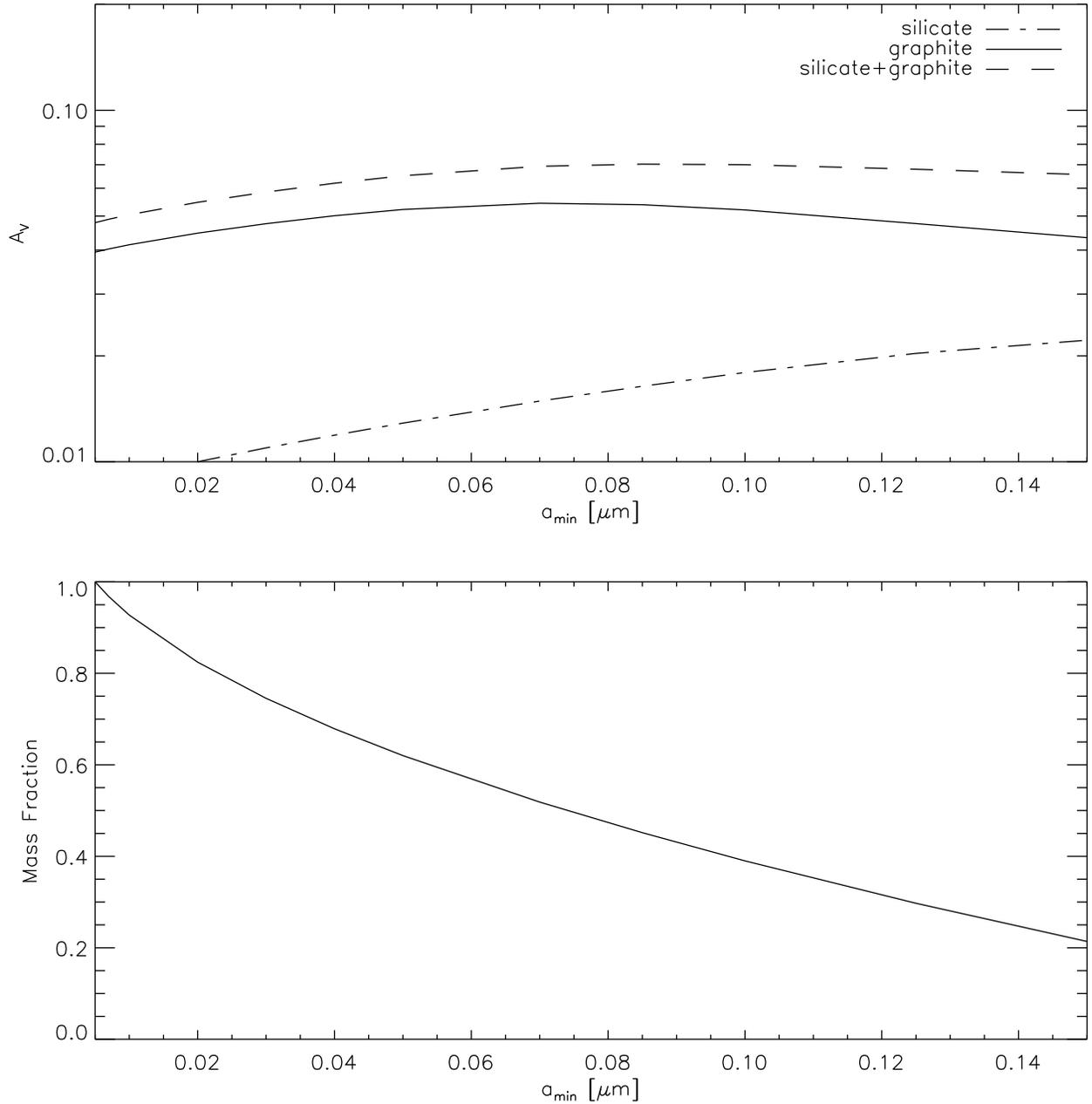}
\caption{{\bf Top:} Extinction to
  $z=0.5$ in $\Omega=0.2$ universe for $\Omega_{dust}^{igm}(z=0.5) =
  10^{-5}$ in graphite (solid), silicate (dashed) and
  graphite+silicate (dot-dashed). {\bf Bottom:} Fraction of full MRN
  distribution contained the distribution with $a_{min} \le a \le 0.25\mic.$}
\label{fig-props}
\end{figure}
\begin{figure}
\plotone{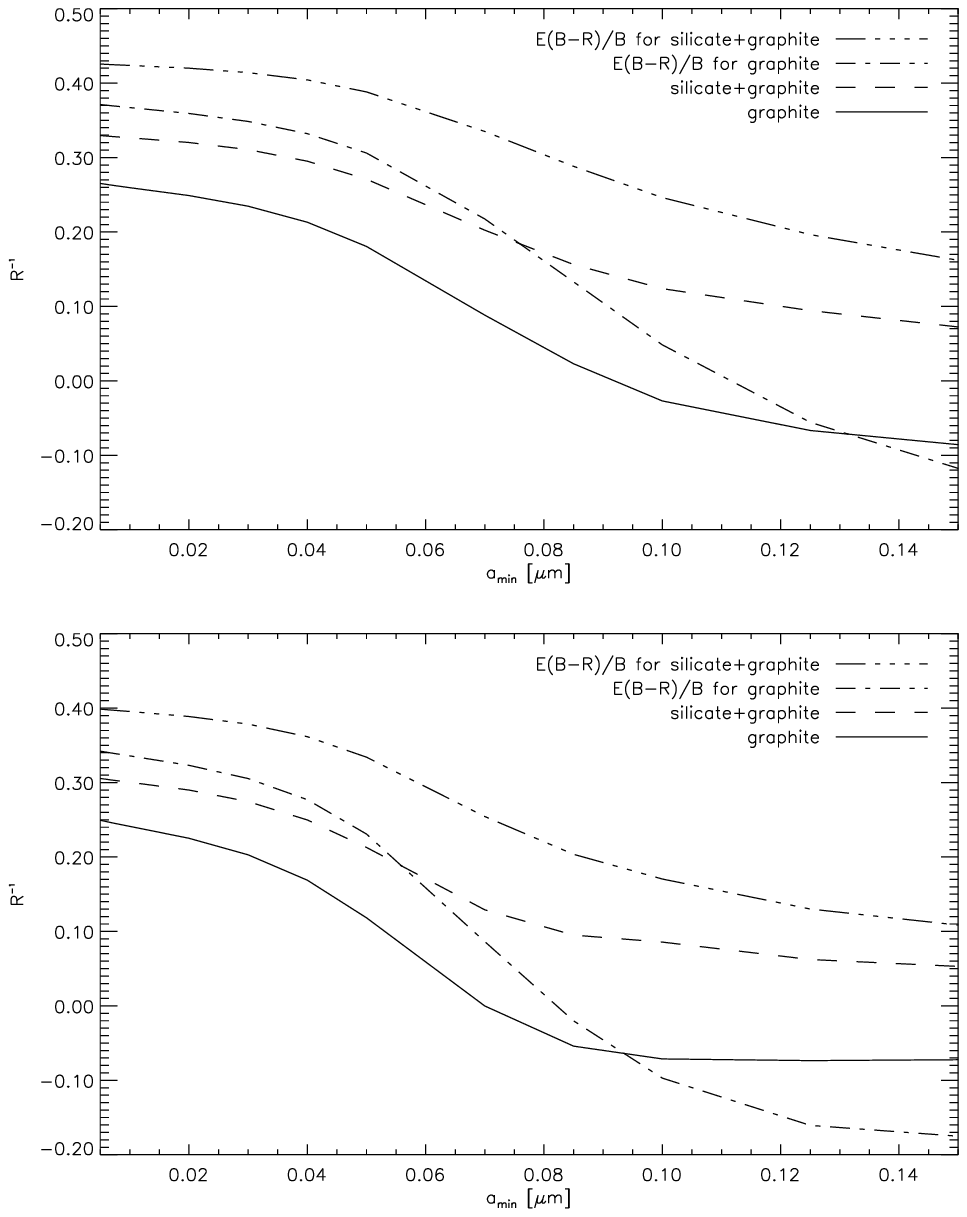}
\caption{{\bf Top:} $E(B-V)/V \equiv R_V^{-1}$ reddening for graphite (solid)
  and silicate+graphite (dashed), for constant comoving dust density
  integrated to $z=0.5$ with $\Omega=0.2$; $E(B-R)/B$ reddening for
  graphite (dot-dashed) and graphite+silicate (triple-dot-dashed). The filters
  refer to rest frame colors at $z=0.5$. {\bf Bottom:} Same, but for
  dust at a single redshift with colors referring to that frame.}
\label{fig-reds}
\end{figure}

Figure~\ref{fig-opacs} shows the
extinction curve for silicate and graphite with the MRN size
distribution over $a_{min} \le a \le a_{max}$ for $a_{min} = 0.005,
a_{max}=0.1$ and $a_{min} = 0.1, a_{max}=0.25$.  These curves use
publicly available extinction data calculated using the method of Laor
\& Draine (1993).  With the very small grains gone, the graphite
absorption curve becomes quite flat out to $\lambda \sim 1\mic.$
Figures~\ref{fig-props} and~\ref{fig-reds} show the extinction,
reddening and mass fraction (relative to the full MRN distribution) of
dust distributions with various value of $a_{min}$.  Curves are given
both for (rest-frame) $E(B-V)/V$ reddening concentrated at one
redshift, and for a cosmological dust distribution (as described
in~\S\ref{sec-snae}).  These show that even in the (more reddening)
integrated extinction, for $a_{min} \ga 0.06\mic$, graphite grains
give very little $(B-V)/V$ reddening. Silicate grains do not become
grey for $a_{min} \la 0.2$, but the combined silicate+graphite
reddening falls by 50\% for $a_{min} \ga 0.09\mic$.  Moreover, this
large change in the reddening behavior of the dust does not require a
large change in the mass: these `grey' dust distributions contain
$40-55\%$ of the mass of the MRN distribution.
\label{sec-dustmod}

\subsection{Other Dust Models}

The above conclusions, based on the assumption that dust is
characterized by the DL model, may not hold for other
dust models.  Mathis \& Whiffen (1989) have proposed that galactic
grains are composites of very small ($a \la 0.005\mic$) silicate,
graphite and amorphous carbon particles.  These composite grains have
a filling factor of $\sim 0.2-1$ and corresponding maximal size $\sim
0.9 - 0.23\mic$.  Sputtering would be effective at destroying all
sizes of low filling-factor composite grains, since both gas drag
(slowing the grains) and sputtering would be much more effective than
in comparably sizes solid spheres.  Also, sputtering might tend to
`cleave' large, filamentary grains into smaller ones. Large
filling-factor particles in this model would be much like the Draine
\& Lee model, although the optical properties of the composite
materials would differ from those of pure graphite or silicate.

Several core-mantle grain models have also been proposed; see e.g.,
Duley, Jones \& Williams (1989) and Li \& Greenberg (1997).  The
latter model assumes a three-component model: large silicate
core-organic refractory mantle dust, very small carbonaceous grains,
and polyaromatic hydrocarbons (PAHs).  The latter two components would
presumably be destroyed as dust leaves the galaxy, leaving the large
core/mantle grains.  Li \& Greenberg take the size distribution of
these grains as Gaussian, strongly dominated by $\sim 0.1\mic$ grains,
with parameters chosen to fit the observed extinction curve.  Such a
distribution would be insignificantly affected by removal of the small
or large-size portions, so intergalactic dust would have properties
exemplified by the large core/mantle grains (these grains will redden
less than the full three-component model, but only slightly).  On the
other hand, it seems that there are good reasons to expect a power-law
grain size distribution (Biermann \& Harwit 1979; Mathis \& Whiffen
1989).  It would be interesting to investigate whether the model of Li
\& Greenberg could accommodate a power law distribution (as they
assume for the PAHs and the very small grains).  The Duley, Jones \&
Williams (1989) model assumes a bimodal grain-size distribution: small
silicate core/graphite coated grains provide UV extinction and the
$0.2175\mic$ bump, whereas an MRN distribution of cylindrical silicate
grains provides extinction in the IR, with grey extinction in UV and
optical.  In this model, intergalactic dust (composed of the large
silicate grains) would be significantly more grey than galactic dust,
assuming that it can escape.

Fractal grains (e.g. Wright 1987) and needles (e.g. A99 and references
therein) provide another possible dust component.  Needles and
platelets have been observed in captured dust (Bradley, Brownlee \&
Veblen 1983), and might explain `very cold' dust in the ISM (Reach et
al. 1995).  As argued in A99 these grains redden very little
(especially if graphitic) and absorb with high efficiency, hence would
be preferentially ejected by radiation pressure.  Along the same
lines, DL grains must be at least somewhat elongated in order to
correctly predict polarization.  Elongated grains are somewhat more
grey than spherical grains of the same mass, giving some additional
support to the general assumption that there is a significant grey
sub-component to interstellar dust.

While a different grain model might predict a different effect of
destroying small grains, it is also true that any viable grain model
must be capable of accommodating values of $R_V \ga 6$, since such
values are in fact observed.  Large grains seem to be a necessary
component of grain models which match the observed extinction laws
(Kim et al. 1994; Zubko, Krelowski \& Wegner 1996, 1998), and a
greater fraction of these large grains in some regions is probably
responsible the high observed values of $R_V$ in those regions.  It
is, then, unlikely that the destruction of very small grains will make
any model {more} reddening, so the assumption of the DL model seems at
least qualitatively safe.  The MRN grain-size distribution is probably
safe for the same reasons, and very likely even conservative, in the
sense that inversions of dust opacity curves into grain-size
distributions tend to lead to more large grains than MRN would predict
(Kim et al.  1994), and the grain-size distribution gleaned from
observations by the Ulysses and Galileo satellites (Frisch et al.
1999) shows many large grains up to $1\mic$ or more in radius.

\label{sec-othermodels}

\section{The supernova results}

Dust of the DL model with $a_{min} \sim 0.1$ would correspond to a
dust survival fraction $f_{esc}f_{igm} \sim 0.4$.  Using
equation~\ref{eq-metdens2}, this gives $\Omega_{dust}^{igm} \approx
(1.5- 4.5) \times 10^{-5}.$ This amount of dust would be quite
important cosmologically.  Measurements of the redshift-magnitude
relation of type Ia supernovae (P99; Riess et al. 1998) show
statistically significant progressive dimming of supernovae which 
has been interpreted as evidence for acceleration in the cosmic
expansion.  This section discusses grey intergalactic dust (as
specified in \S\ref{sec-dustmod}) in the context of the these
observations.  Only the DL grain model is considered here.

\begin{figure}
\plotone{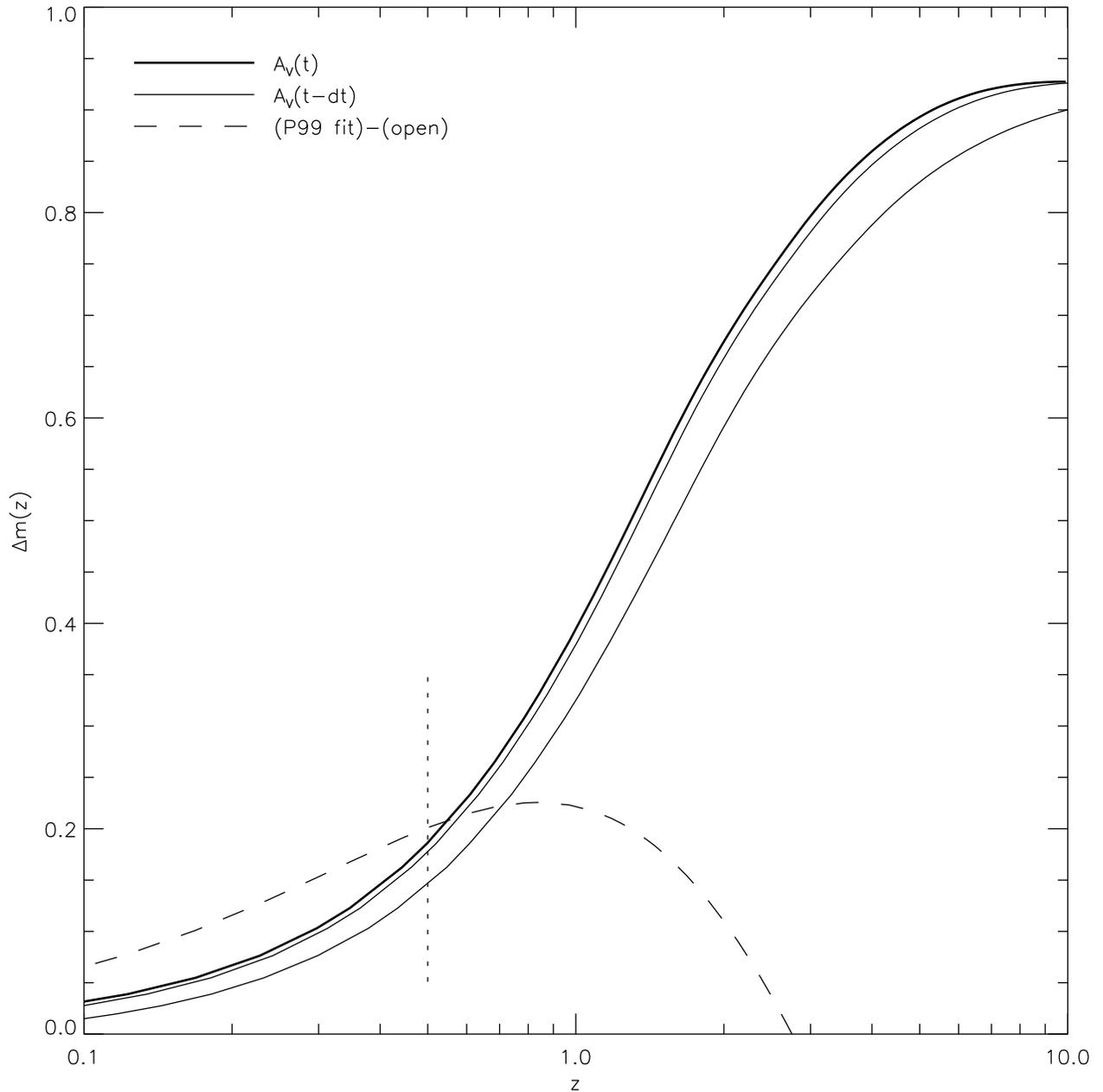}
\caption{Deviations from an open universe:
  Optical depth to redshift $z$ in ($z=0.5$ rest-frame) $V$-band
  (solid, dark). Deviation of the P99
  fit $\Omega_\Lambda=0.72, \Omega=0.28$ from a $\Omega=0.2$ open
  cosmology (dashed).  Solid, light lines indicated optical depth due to dust
  formed $dt=200$ Myr and $dt=1000$ Myr earlier than indicated by dark
  solid line.  The dotted line signifies the `intrinsic' dispersion in
  observed supernova magnitudes about the fit.  The assumed dust density is
  proportional to the integrated (Madau 1999) SFR (see
  figure~\ref{fig-sfr}), normalized to $\Omega_{dust}^{igm}(z=0)=4.5\times
  10^{-5}$.}
\label{fig-depth}
\end{figure}

Both supernova groups find that after calibration using a low-$z$
sample, the supernova at $z\sim0.5$ have magnitudes indicative of
acceleration in the cosmic expansion.  The best fit (for a flat
cosmology with cosmological constant) of P99 is $\Omega=0.28,
\Omega_\Lambda=0.72$; the results of R99 are similar.  The necessary
extinction to account for these results in an $\Omega=0.2$ open
universe (see Fig.~\ref{fig-depth}) is $A_V(z=0.5) \approx 0.15-0.2$
mag.\footnote{ $A_V \approx 0.2$~mag accounts for all of the effect,
  but $A_V \approx 0.15$~mag puts an $\Omega=0.2$ open universe within
  the stated $1-\sigma$ contour.}  An $\Omega=1$ universe requires
$A_V(z=0.5) \approx 0.4$ mag.

Riess et al. argue that grey extinction would cause too much
dispersion in the supernova magnitudes to be compatible with their
observations if the dust is confined to spiral galaxies.  Perlmutter
et al. derive from their data an intrinsic dispersion $\Delta$ at
$z\sim 0.5$ almost identical to that at $z\sim 0.05$:
$$\Delta(z\sim0.05) = 0.154\pm0.04,\ \ \ \Delta(z\sim0.5) =
0.157\pm0.025.$$
This suggests that the processes dominating the
intrinsic dispersion do not change significantly in magnitude from low
to high redshift.  However, note that -- assuming that the errors as
well as the dispersions add in quadrature -- the amount of {
  additional} dispersion $\Delta_{add}$ at high $z$ formally allowed
within the stated errors of is $\Delta_{add} \la 0.13$ mag.  This does
not include any systematic errors in the estimation of the intrinsic
dispersion.

P99 also investigate the mean color difference between
the high- and low-$z$ samples, finding
$\langle E(B-V)\rangle_{z\sim 0.05} = 0.033\pm0.014$ and $\langle
E(B-V)\rangle_{z\sim 0.5} = 0.035\pm0.022.$
Again, this suggests that
a systematic effect (in color) is not large, but nevertheless the
errors allow a color difference of 
$$\langle E(B-V)\rangle_{z\sim 0.5} - \langle E(B-V)\rangle_{z\sim
  0.05} \la 0.03\,{\rm mag}.$$
In addition, this comparison is subject
to a systematic uncertainty of $\approx 0.03$ mag resulting from the
conversion of (observed) $R$ and $I$ magnitudes into rest-frame $B$
and $V$ magnitudes.

To place tighter constraints on systematic reddening, Perlmutter et
al. construct an artificially blue subsample of the high-$z$ points
which is unlikely to be redder (in the mean) than the low-$z$ sample.
The change in fitting that this elimination produces then gives an
indication of systematic extinction by reddening dust.

Perlmutter et al. use this method to place a strong constraint of
$\delta A_V \la 0.025$ mag on the effect of any extinction which (a)
exists at high $z$ but not at low $z$, (b) dominates the dispersion of
both the color and extinction, (c) has a reddening-extinction relation
$R_V$ up to twice that of the Galaxy and (d) occurs in a flat
universe.  Assumption (b), unstated in P99, is crucial but seems
unfounded.  The limit on a systematic increase in dispersion indicates
that systematic extinction must be a sub-dominant component of the
total computed `intrinsic' dispersion in magnitude; this holds also
for dispersion in color.  In this case, removing the reddest
supernovae will not preferentially remove more obscured supernovae,
even if dust accounts for the whole effect at high $z$.  Thus the
stated (more stringent) limits on systematic reddening do not apply,
{as long as} the dispersion in brightness and/or color is dominated by
factors other than extinction. Furthermore,
if the assumption of flatness is dropped, the elimination of the seven
reddest supernovae actually changes the fit considerably, in the
direction of an open universe (P98, Table 3, Figure 5c).  The shift
corresponds to $\delta A_V \approx 0.07$ mag at z=0.5.


The required extinction and the limits on reddening and dispersion can
now be compared to that expected from intergalactic dust.
Figure~\ref{fig-reds} shows the reddening $R_V$ for graphite and
silicate dust of the DL model, assuming an MRN distribution over
$a_{min} \le a \le 0.25\mic.$ The extinction in Figure~\ref{fig-props}
is integrated to $z=0.5$ for an $\Omega=0.2$ universe, assuming a
constant comoving dust density of $\Omega_{dust}^{igm}=10^{-5}$ in
each component. The results show that, for example, a distribution
with $a_{min} = 0.1\mic$, $\Omega_{dust}^{igm}(z=0.5) \approx 4 \times
10^{-5}$ (total) and equal mass density of silicate and graphite
grains (approximately the ratio derived by Draine \& Lee) provides
sufficient extinction to account for the type Ia supernova results.
The induced reddening is $0.025$ mag, comparable to the allowed
reddening due to {either} random {or} systematic errors.  Most
of this reddening is provided by the silicate grains, so the 1:1 ratio
is conservative; the real ratio should be biased toward the less
efficiently destroyed (Draine \& Salpeter 1979a) and possibly more
efficiently ejected (assuming radiation pressure expulsion) graphite
grains.  The large grains contain $\sim 40\%$ of the full MRN
distribution.  Larger values of $a_{min}$ provide less reddening but
values of $a_{min} \ga 0.15$ probably contain too little mass to be
viable in explaining the supernova data.  Graphite grains alone (if
silicate grains were preferentially destroyed) with $a_{min} \ga 0.06$
(giving $f_{esc}f_{igm} \la 0.6$) and $\Omega_{dust}^{igm} \sim 3
\times 10^{-5}$ would produce similar effects.

The amount of dispersion induced by the dust is very important but can
be estimated only roughly.  Assuming that the dust is uniformly
distributed in randomly placed spheres of radius $R$ with number
density $n$, the dispersion $\Delta$ is given approximately by
$\Delta/A_V(z=0.5) \approx N^{-1/2}$, where $N$ is the number of
spheres intersected by a typical path, and can be written $N \simeq
n\pi R^2D$, where $D\approx 2400h_{65}^{-1}\,$Mpc is the distance to
$z=0.5$ in an $\Omega=0.2$ universe.
Now consider galaxies with $n = 0.008h_{65}^3(1+z)^3\,{\rm
  Mpc^{-3}}$ (Lin et al. 1996).  For $N^{1/2} \ga 1$ this implies $R
\ga 70h_{65}^{-1}[(1+z)/1.5]^{-3/2}\,{\rm kpc}$.  Escape velocities from
spirals are $\ga 250\,{\rm km\,s^{-1}}$, so the dispersion in
integrated optical depth due to dust ejected by radiation pressure or
winds and traveling away from the disk for time $\tau_{esc}$ is
\begin{equation}
\Delta \sim A_V h_{65}^{-1}\left({\tau_{esc} \over 270\,{\rm
      Myr}}\right)^{-1}\left({v_{dust}\over 250\,{\rm
      km\,s^{-1}}}\right)^{-1}.
\end{equation}
Figure~\ref{fig-depth} shows the
(small) difference in optical depth between the total dust
distribution and the dust which has existed for $> 200$ Myr and $> 1$
Gyr, demonstrating that the dispersion induced by grey dust created
but not yet sufficiently dispersed would be small.  Large-scale correlations
between galaxies are unlikely to be important in this analysis.  The
dispersion in density on 8 Mpc scales is $\sim 1$, but $\sim 300$ such
domains lie between here and $z\sim 0.5$, leading to $\Delta \sim
0.2/\sqrt{300} = 0.01$.  

The best way to estimate dispersion (currently underway) would
probably be to measure the optical depth through random lines of sight
piercing a high-resolution cosmology simulation which either tracks
metals (Cen \& Ostriker 1999b; Gnedin 1998) or allows some
prescription relating gas density to dust density in the IGM.  Short
of this, I note that Cen \& Ostriker's simulations show that $\delta
\sim 100$ is characteristic of the bulk of the metal-rich gas.  This
corresponds to $R \ga 670h_{65}^{-1}(1+z)^{-1}$ kpc for the spheres
considered above.  Fairly uniform regions of this size and number
density would give little dispersion.

If $R \la 70$~kpc, most extinction takes place in a small number of
clumps, in particular in the halo of the supernova host galaxy.  For
dust of density $\Omega_{dust} = \chi\times 10^{-5}$ uniformly
distributed in radius $R=\xi \times 100\,{\rm kpc}$ halos of galaxies
with $n = 0.008h_{65}^3(1+z)^3\,{\rm Mpc^{-3}}$, the extinction to the
galaxy center is
\begin{eqnarray}
A_V &=& {1.086\kappa_V\Omega_{dust}\rho_c\over {4\over 3}\pi R^2n} \\ \nonumber
&=& 0.04 \left({\kappa_V \over 5 \times 10^{4}\,{\rm cm^2\,g^{-1}}}\right)
h_{65}^{-1}\chi\xi^{-2}, \\ \nonumber
\label{eq-galopt}
\end{eqnarray} 
which gives the required extinction for $\xi \approx 1$ and $\chi
\approx 4.$ As long as $\xi \ga 0.1$ there will not be large
dispersion due to different galactic radii in the supernovae, but
inhomogeneities in the dust distribution in the halo could be
important.  The supernova results could could be explained by such
halos with a smaller $\Omega_{dust}$, but this would require that the
halos are somewhat larger at low $z$, and that all halos at $z\sim
0.5$ have similar column density through them. This scenario does not
seem to be as natural an explanation as a more uniform dust
distribution, but it remains a possibility.

Leaving aside these uncertainties, the essential result of my
calculation is that a truncated-MRN distribution of Draine \& Lee dust
with $a_{min} \ga 0.1$, uniformly distributed and with
$\Omega_{dust}^{igm}(z=0.5) \simeq 4\times 10^{-5}$ can account for the
supernova dimming in an $\Omega=0.2$ universe without excessive
reddening.  Whether the induced dispersion in extinction is too large
depends crucially on the (uncertain) distance to which ejected dust
can escape from the galaxies in which it forms, and on the clumpiness
of the resulting distribution.

\label{sec-snae}

\section{Cosmic backgrounds}
 
An intergalactic dust distribution of the magnitude required to
explain the supernova results would have other cosmological
implications.  For instance, any intergalactic dust component will
absorb energy from the optical/UV background and re-emit the energy in
the FIR/microwave.  Calculation of the evolution of the cosmic mean
density field shows (Aguirre \& Haiman, in preparation) that the dust
considered in this paper will not lead to measurable CMB spectral
distortions,\footnote{Intergalactic dust in the calculations of Loeb
  \& Haiman (1997) and Ferrara et al. (1999) has low temperature, but
  both papers assume a UV/optical background lower than recent
  detections indicate.}  but instead adds (significantly) to the CIB.
The FIRAS distortion limits can, however, limit dust types with high
FIR emissivity, since these have lower equilibrium temperatures.

Note also that while the SFR estimates correct for dust extinction
(e.g. Madau, Pozetti \& Dickinson 1998; Pettini 1999 and references
therein), these corrections would not account for intergalactic dust.
Intergalactic absorption of $\sim 0.1 - 1$ mag at $z \sim 0.5-5$ (see
fig.~\ref{fig-depth}) would imply an SFR -- and hence metal density --
a factor of $\sim 2$ higher than that given in~\S\ref{sec-metals}.

\section{Testing for intergalactic dust}

Future observations of supernovae can investigate the importance of
intergalactic dust in two ways. First, accurate high-$z$ observations
in rest-frame $R$-band or longer wavelengths should reveal the dust
with properties of the model developed here.  As shown in
fig.~\ref{fig-reds}, the $E(B-R)/B$ reddening to $z=0.5$ is $\approx
0.25$ for both pure graphite dust (with $a_{min} = 0.06\mic$) or for
silicate+graphite (with $a_{min}=0.1\mic$). Therefore high-$z$
supernova should have $E(B-R)$ values 0.05 mag higher than the low-$z$
sample.  The effect in $B-I$ would be even stronger. Non-spherical
grains (of which the needles of A99 are an example) could provide grey
opacity into $R$-band and beyond, but are constrained limits on
far-infrared/microwave emission (Aguirre \& Haiman, in preparation).


Figure~\ref{fig-depth} includes the deviation of the fit of R99
from the $\Omega=0.2$ model (dashed line).  The difference
between this line and the dust optical depth then indicates the
deviation at high $z$ of the dust model from the cosmological constant
($\Omega_\Lambda = 0.72$) model; at $z=1.5$ the difference is $\simeq
0.2-0.3$ mag.  This possible difference should be testable once a
significant number of $z > 1$ supernovae are measured.  A large number
of supernovae is important both because the dispersion is comparable
to the effect being measured, and because it is crucial to be sure
that the $z>1$ sample is statistically {complete} -- a worry that
is less significant at $z\sim0.5$ since the claimed effect is of
dimming rather than brightening.

An intergalactic dust distribution would also dim other distant
objects.  Dust has been proposed several times as an explanation for
the dropoff in quasar number counts for $z\ga 3$ (e.g. Ostriker \&
Heisler 1984; Wright 1986).  While this point is still controversial,
observation of the dropoff in radio-selected quasars (Shaver et al.
1996) shows this explanation to be unlikely.  The models of this paper
would not predict the dropoff (unless substantial population III dust
exists also) since $A_V(z) \la 1$ for all $z$. Galaxy counts, in
contrast to quasar studies, show an excess in $B$-band counts at
high $z$ as compared to some evolution models (see Shimasaku \&
Fukugita 1998 for a summary).  For $B$-band observations probing
$z\sim 0.5\pm0.3$, this `excess' amounts to a deviation of up to $\sim
1$ mag from the predicted curve.  The dust extinction of $\la 0.4$ mag
for $z \la 1$ would enlarge this discrepancy, but not greatly.  This
holds for $I$ and $K$-band surveys as well.  Because of uncertainties
in the models at high $z$, Shimasaku \& Fukugita are reluctant to draw
conclusions about cosmological parameters from the galaxy counts.  But
with rapid progress in the field this may soon become a useful test of
the dust model.

Several claims of dust detection in clusters have been made, but
remain controversial.  The model of A99 predicted that cluster dust
should be more grey than galactic dust; the arguments of this paper
reinforce this prediction.  The intracluster gas destroys dust
efficiently; since sputtering destroys small grains more efficiently
than large grains, it is likely that whatever grains survive in
clusters are large, and thus supply grey opacity (this may help
explain the controversy surrounding the existence on intracluster
dust).  Determination of the extinction curve of dust in clusters (if
dust indeed exists) would be an important test of the key idea of the
intergalactic dust model, although grey dust in clusters would not
necessarily imply that dust in the diffuse IGM is also grey.

Dust confined to galactic halos might also be detectable by its IR
emission or by studies of objects seen through the halo.  Zaritsky
(1994) has presented preliminary evidence for halo dust in NGC 2835
and NGC 3521 at $\sim 60\,$kpc using $B-I$ reddening; if confirmed
this would be an important dust component, as the inferred mass is
large.

\section{Conclusions}

As discussed in \S\ref{sec-metals}, fairly strong arguments suggest
that the universe currently has $\Omega_Z \approx (1.5-3) \times
10^{-4}$ in metals.  This metal cannot all be contained in the stars and
gas in known galaxies, so unless metals are well hidden in unobserved
galaxies, a metal density $\Omega_Z^{igm} \approx (0.75-2.25) \times 10^{-4}$
should exist in the more uniform IGM or in extended halos. Moreover,
most of this metal was probably in place by $z\sim0.5$.

Some fraction $d_m$ of this metal must be in dust.  There are several
good reasons to expect that the grain size distribution of this dust should
be different than for dust in galaxies: (1) Ejection by radiation
pressure favors higher opacity, less reddening grains (A99). Within
the DL model, this means large grains are preferentially ejected. (2)
As dust leaves galaxies (where it is almost certainly created), small
grains are preferentially destroyed by sputtering.  This preferential
destruction occurs also in the IGM, but the efficiency is rather
uncertain. (3) Small grains are generally assumed to form from the
shattering of larger grains.  This shattering will not occur in the
IGM due to the low densities, nor can small grains grow from the
vapor.

The principal assumption of this paper (supported by what detailed
calculations are available) is that very small dust grains leaving the
galaxy -- comprising a fraction $(1-f)$ of the total dust density --
are removed from the grain-size distribution, while large grains are
not.  Dust as modeled by DL, with grains of radius ($\la 0.1\mic$)
removed ($f \approx 0.4$), reddens very little yet has a visual
opacity $\kappa \approx 5\times 10^4\,{\rm cm^2\,g^{-1}}$.  Uniformly
distributed dust of constant comoving density in an $\Omega=0.2$ open
universe provides an extinction to $z=0.5$ of
\begin{equation}
A_V \simeq 0.15h_{65}\left({\Omega_{dust}^{igm} \over 4\times 10^{-5}}\right)
\left({\kappa\over 5\times 10^4\,{\rm
      cm^2\,g^{-1}}}\right),
\end{equation}
hence it can
account for the dimming of type Ia supernovae at $z\sim 0.5$ in a way
fully consistent with observations.  The fact that the expected dust
density of $0.5 f \Omega_Z^{igm}  \simeq (0.25-2.25)\times 10^{-4} \times
(0.5) \times (0.4) = (1.5-4.5) \times 10^{-5}$ is so close to
the density required is very interesting.
According to these estimates (see also~\S\ref{sec-metals} and \S\ref{sec-dustdens}) it is
possible, but rather unlikely, that there is sufficient dust
($\Omega_{dust}^{igm}(z=0.5) \sim 9 \times 10^{-5}$) to allow
compatibility with a closed, matter-dominated universe.

The arguments of this paper show that there may be an intergalactic
dust component which is important yet has evaded earlier attempts at
its discovery; from this point of view the supernova observations are
telling us not about cosmic acceleration, but about cosmic opacity.
If the arguments of this paper are substantially correct, an
intergalactic dust distribution of {some} magnitude is inevitable, so
it is worth pointing out the chief caveats (which maintain rough
consistency with other observations) under which the obscuration would
not be sufficient to to account for the supernova dimming:
\begin{itemize}
\item{Both the SFR argument and the cluster enrichment argument may
    predict cosmic metallicity several times the true value.  In this
    case most metal could be locked in galaxies.  Taking
    this view, however, probably also requires one to assume that cluster
    galaxies both have a different IMF and eject metals more
    efficiently than field galaxies do.}
\item{A large population of unobserved but compact objects such a low
    surface brightness galaxies could contain a large fraction of the
    cosmic metallicity.  This would, however, also fail to explain the
    cluster observations.}
\item{The Draine \& Lee model may not accurately characterize dust, so
    the effect of small grain destruction on the opacity curve may be
    significantly different than assumed here.  But the alternative
    grain model would then probably have to exclude the grey
    sub-component which most current grain models include, yet still
    explain the values $R_V \ga 6$ observed in some regions.}

\item{Dust of all sizes may be efficiently destroyed in the diffuse
    IGM or as it leaves galaxies.  While total grain destruction does
    not seem to be quantitatively supported in the DL model, it is
    certainly plausible, and would probably be the case if grains are
    very fluffy, high filling-factor composites.}
\end{itemize}

It is unlikely that the issue can be cleanly decided on the basis of
the above points or by the current supernova data (i.e. limits on
dispersion or reddening); on the other hand, future supernova results
can rule decisively.  Statistically robust deviation of the
magnitude-redshift curve of $z>1$ supernovae from the dust prediction
would argue strongly for the interpretation of the dimming as cosmic
acceleration and for the relative unimportance of grey intergalactic
dust.  Clear evidence of, for example, systematic (rest frame) $B-R$
reddening would argue strongly for dust.

If future observations show that intergalactic dust is indeed
important, observations of type Ia supernovae will be seen to have not
only determined the deceleration parameter (once dust is accounted
for), but to have discovered a component of the universe that will
have important implications for many future observations at high
redshift.

\acknowledgements

I thank David Layzer, George Field, Andrea Ferrara, Zoltan Haiman,
Chris Kochanek, Pat Thaddeus, Bob Kirshner, Eliot Quataert and Saurabh
Jha for helpful discussions.  Communications from Ari Laor, Anthony
Jones, Tom Yorke, and an an anonymous referee were helpful and
appreciated. I thank also Bruce Draine for making publicly available
the dust data used, and I am grateful to Bill Press for financial
support.  This work was supported in part by the National Science
Foundation grant no. PHY-9507695.

\newpage

\end{document}